\documentclass{appolb}
\usepackage{epsfig}
\begin{document}
\eqsec 
\title{COMPLEXITY CHARACTERIZAZION OF DYNAMICAL SYSTEMS THROUGH  
                         PREDICTABILITY
\thanks{Presented at XV Mariam Smoluchowski Symposium on Statistical 
Physics: Zakopane Poland Sept. 7-12, 2002}
}
\author{Fabio Cecconi 
\and Massimo Falcioni
\and
Angelo Vulpiani
\address{Dipartimento di Fisica Universit\`a di Roma ''La Sapienza''\\
 INFM, Unit\`a di Roma1 and SMC Center,\\ 
 P.le Aldo Moro 2, 00185 Roma, Italy}
}
\maketitle
\begin{abstract}
Some aspects of the predictability problem in dynamical systems are
reviewed. The deep relation among Lyapunov exponents, Kolmogorov-Sinai
entropy, Shannon entropy and algorithmic complexity is discussed. In
particular, we emphasize how a characterization of the
unpredictability of a system gives a measure of its complexity.  A
special attention is devoted to finite-resolution
effects on predictability, which can be accounted with suitable
generalization of the standard indicators. The problems involved in
systems with intrinsic randomness is discussed, with emphasis on the
important problems of distinguishing chaos from noise and of modeling
the system.
\end{abstract}

\PACS{PACS 45.05.+x, 05.45.-a}

\hfill\break
  
\noindent {\it All the simple systems are simple in the same way, each
complex system has its own complexity} (freely inspired by {\it Anna
Karenina} by Lev N. Tolstoy)

\section{Introduction}
The possibility to predict future states of a system stands at the
foundations of scientific knowledge with an obvious relevance both
from a conceptual and applicative point of view.  The perfect
knowledge of the evolution law of a system may induce the conclusion
that this aim could be attained.  This classical deterministic point
of view was claimed by Laplace \cite{L1814}: once the
evolution laws of the system are known, the state at a certain time
$t_0$ completely determines the subsequent states for every time $t >
t_0$.  However it is well established now that in some systems, full
predictability cannot be accomplished in practice because of the
unavoidable uncertainty in the initial conditions. Indeed, as already
stated by Poincar\'e, long-time predictions are reliable only when the
evolution law does not amplify the initial uncertainty too rapidly.
Therefore, from the point of view of predictability, we need to know
how an error on the initial state of the system grows in time.  In
systems with great sensitive dependence on initial conditions
(deterministic chaotic systems) errors grows exponentially fast in
time, limiting the ability to predict the future states.

A branch of the theory of dynamical systems has been developed with
the aim of formalizing and characterizing the sensitivity to initial
conditions. The Lyapunov exponent and the Kolmogorov-Sinai entropy are
the two main indicators for measuring the rate of error growth and
information production during a deterministic system evolution.  A
complementary approach has been developed in the context of
information theory, data compression and algorithmic complexity theory
and it is rather clear that the latter point of view is
closely related to the dynamical systems one.  If a system is chaotic,
then its predictability is limited up to a time which is related to
the first Lyapunov exponent, and the time sequence by which we encode
one of its chaotic trajectories cannot be compressed by an arbitrary
factor, i.e. is algorithmically complex.  On the contrary, the coding
of a regular trajectory can be easily compressed (e.g., for a periodic
trajectory it is sufficient to have the sequence for a period) so it
is ``simple''.

In this paper we will discuss how unpredictability and algorithmic
complexity are closely related and how information and chaos theory
complete each other in giving a general understanding of complexity in
dynamical processes.  In particular, we shall consider the extension
of this approach, nowadays well established in the context of low
dimensional systems and for asymptotic regimes, to high dimensional
systems with attention to situations far from asymptotic (i.e. finite
time and finite observational resolution) \cite{Boffy}.

\section{Two points of view}

\subsection{Dynamical systems approach:
Characteristic Lyapunov exponents}

The characteristic Lyapunov exponents are somehow an extension of the
linear stability analysis to the case of aperiodic motions.  Roughly
speaking, they measure the typical rate of exponential divergence of
nearby trajectories and, thus, contain information on the growing rate
of a very small error on the initial state of a system.

Consider a dynamical system with an evolution law given, e.g., by the
differential equation
\begin{equation}
\label{eq:1-1}
{ d {\bf x} \over dt } =  {\bf F} ({\bf x}) \; ;
\end{equation}
we assume that ${\bf F}$ is smooth enough that the evolution is
well-defined for time intervals of arbitrary extension, and that the
motion occurs in a bounded region of the phase space.  We intend to
study the separation between two trajectories, ${\bf x}(t)$ and ${\bf
x}'(t)$, starting from two close initial conditions, ${\bf x}(0)$ and
${\bf x}'(0) = {\bf x}(0) + \delta {\bf x}(0)$, respectively.

As long as the difference between the trajectories, $\delta {\bf x}(t)
= {\bf x}'(t) - {\bf x}(t)$, remains small (infinitesimal, strictly
speaking), it can be regarded as a vector, ${\bf z}(t)$, in the
tangent space.  The time evolution of ${\bf z}(t)$ is given by the
linearized differential equations:
\begin{equation}
\label{eq:1-3}
   {d z_i(t)\over d t}= \sum_{j=1}^d\,
   \left. {\partial F_i  \over \partial x_j} 
   \right|_{ {\bf x}(t)}\, z_j(t) \; . 
\end{equation}
Under rather general hypothesis, Oseledec \cite{O68} proved that for
almost all initial conditions ${\bf x}(0)$, there exists an
orthonormal basis $\lbrace {\bf e}_i \rbrace$ in the tangent space
such that, for large times,
\begin{equation}
\label{eq:1-5}
{\bf z}(t) = \sum _{i=1}^{d} c_i {\bf e}_i e^{\lambda_i \, t} \,, 
\end{equation}
where the coefficients $\{c_i\}$ depend on ${\bf z}(0)$.  The
exponents $\lambda_1 \ge \lambda_2 \ge\cdots\ge \lambda_d$ are called
{\it characteristic Lyapunov exponents} (LEs). If the dynamical system
has an ergodic invariant measure, the spectrum of LEs $\lbrace
\lambda_i \rbrace $ does not depend on the initial condition, except
for a set of measure zero with respect to the natural invariant
measure.

Equation~(\ref{eq:1-5}) describes how a $d$-dimensional spherical
region of the phase space, with radius $\epsilon$ centered in ${\bf
x}(0)$, deforms, with time, into an ellipsoid of semi-axes $\epsilon
_i (t) = \epsilon \exp (\lambda _it)$, directed along the $ {\bf e}_i
$ vectors.  Furthermore, for a generic small perturbation $\delta {\bf
x}(0)$, the distance between the reference and the perturbed
trajectory behaves as 
$$|\delta {\bf x} (t)|\sim |\delta {\bf x} (0)|
\, e^{\lambda_1\, t} \, \left[ 1 +
O\left(\exp{-(\lambda_1-\lambda_2)t}\right) \right].
$$ 
If $\lambda_1 > 0$ we have a rapid (exponential) amplification of an
error on the initial condition. In such a case, the system is chaotic
and, {\it de facto}, unpredictable on the long times. Indeed, if the
initial error amounts to $\delta_0 = |\delta {\bf x} (0)|$, and we
purpose to predict the states of the system with a certain tolerance
$\Delta$ (not too large), then the prediction is reliable just up to a
{\it predictability time} given by
\begin{equation}
\label{Tpred}
T_p \sim {1 \over \lambda_1} \ln \left({\Delta \over \delta_0}\right)\, .
\end{equation}
This equation shows that $T_p$ is basically determined by the largest
Lyapunov exponent, since its dependence on $\delta_0$ and $\Delta$ is
logarithmically weak. Because of its preeminent role, $\lambda_1 $ is
often referred as ``the Lyapunov exponent'', and denoted by $\lambda$.

\subsection{Information based approach}

In experimental investigations of physical processes, the access to a
system occurs only through a measuring device which produces a time
record of a certain observable, i.e. a sequence of data. In this
regard a system, whether or not chaotic, generates messages and may be
regarded as a source of information whose properties can be analysed
through the tools of information theory.

The characterization of the information contained in a sequence can be
approached in two very different frameworks.  The first considers a
specific message (sequence) as belonging to the ensemble of all the
messages that can be emitted by a source, and defines an average
information content by means of the average compressibility properties
of the ensemble \cite{S48}.  The second considers the problem of
characterizing the universal compressibility (i.e. ensemble
independent) of a specific sequence and concerns the theory of
algorithmic complexity and algorithmic information theory
\cite{K65,S64}.  For the sake of self-consistency we briefly recall
the concepts and ideas about the Shannon entropy\cite{S48}, that is
the basis of whole information theory

\subsubsection{Shannon entropy}
Consider a source that can output $m$ different symbols; denote with
$s_t$ the symbol emitted by the source at time $t$ and with $P(C_N)$
the probability that a given word $C_N=(s_1,s_2,\dots,s_N)$, of length
$N$, is emitted $P(C_N)=P(s_1,s_2,\dots,s_N)$.  We assume that the
source is stationary, so that, for the sequences $\{s_t\}$, the time
translation invariance holds:
$P(s_1,\dots,s_N)=P(s_{t+1},\dots,s_{t+N})$.  We introduce the
$N$-block entropies
\begin{equation}
H_N=-\sum_{\{C_N\}} P(C_N)\ln P(C_N)\,,
\label{eq:block}
\end{equation} 
for stationary sources the limit
\begin{equation}
\lim_{N \to \infty} {H_N \over N} = h_{Sh}
\label{eq:shannon}
\end{equation}
exists and defines the Shannon entropy $h_{Sh}$ which quantifies the
richness (or ``complexity'') of the source emitting the sequence.
This can be precisely expressed by the first theorem of
Shannon-McMillan \cite{K57} that applies to stationary ergodic
sources: The ensemble of $N$-long subsequences, when $N$ is large
enough, can be partitioned in two classes, $\Omega_1(N)$ and
$\Omega_0(N)$ such that all the words $C_N \in \Omega_1(N)$ have the
same probability $P(C_N)\sim \exp(-Nh_{Sh})$ and
\begin{equation}
\sum_{C_N \in \Omega_1(N)} P(C_N) \to 1 \qquad
\mbox{while}
\sum_{C_N \in \Omega_0(N)} P(C_N) \to 0 
\qquad {\mbox{ for }} N\to \infty
\end{equation}
The meaning of this theorem is the following. An $m$-states process
admits, in principle, $m^N$ possible sequences of length $N$.  However
the number of typical sequences, $N_{eff}(N)$, effectively observable
(i.e. those belonging to $\Omega_1(N)$) is
\begin{equation}
N_{eff}(N) \sim \exp(Nh_{Sh})\,.
\label{eq:wordstypical}
\end{equation}
Note that $N_{eff}\ll m^N$ if $h_{Sh}<\ln m$. The entropy per symbol,
$h_{Sh}$, is a property of the source.  However, because of the
ergodicity $h_{Sh}$ can be obtained by analyzing just one single
sequence in the ensemble of the typical ones, and it can also be
viewed as a property of each typical sequence.

In information theory, expression~(\ref{eq:wordstypical}) is somehow
the equivalent of the Boltzmann equation in statistical
thermodynamics: $S \propto \ln W$, being $W$ the number of possible
microscopic configurations and $S$ the thermodynamic entropy, this
justifies the name ``entropy'' for $h_{Sh}$.

The relevance of the Shannon entropy in information theory is given by
the fact that $h_{Sh}$ sets the maximum compression rate of a sequence
$\{s_1,s_2,s_3, \dots\}$. Indeed a theorem of Shannon states that, if
the length $T$ of a sequence is large enough, there exists no other
sequence (always using $m$ symbols), from which it is possible to
reconstruct the original one, whose length is smaller than $(h_{Sh} /
\ln m)T$ \cite{S48}.  In other words, $h_{Sh}/\ln m$ represents the
maximum allowed compression rate. The relation between Shannon
entropy and data compression problems is well illustrated by
considering the optimal coding (Shannon-Fano) to map ${\cal N}$
objects (e.g. the $N$-words $C_N$) into sequences of binary digits
$(0,1)$ \cite{N1}.  Denoting with $\ell_N$ the binary length of the
sequence specifying $C_N$, we have
\begin{equation}
\lim_{N \to \infty} {\langle \ell_N \rangle \over N}
={h_{Sh} \over \ln 2} \,,
\label{eq:fano}
\end{equation}
i.e., in a good coding, the mean length of a $N$-word is equal to $N$
times the Shannon entropy, apart from a multiplicative factor, since
in the definition (\ref{eq:shannon}) of $h_{Sh}$ we used
the natural logarithm and here we want to work with a two symbol code.

\subsubsection{The Kolmogorov-Sinai entropy}

After the introduction of the Shannon entropy we can easily define the
Kolmogorov-Sinai entropy which is the analogous measure of complexity
applied to dynamical systems.  Consider a trajectory, ${\bf x}(t)$,
generated by a deterministic system, sampled at the times $t_j = j \,
\tau$, with $j=1, 2,3, \dots$. Perform a finite partition ${\cal A}$
of the phase space, with the finite number of symbols $\lbrace s
\rbrace_{{\cal A}}$ enumerating the cells of the partition. The
time-discretized trajectory ${\bf x}(t_j)$ determines a sequence
$\lbrace s(1), s(2), s(3), \dots \rbrace$, whose meaning is clear: at
the time $t_j$ the trajectory is in the cell labeled by $s(j)$.  To
each subsequence of length $N\cdot\tau$ one can associate a word of
length $N$: $ W^N_j ({\cal A})= \left( s(j), s(j+1) , \dots,
s(j+(N-1)) \right) $. If the system is ergodic, as we suppose, from
the frequencies of the words one obtains the probabilities by which 
the block entropies $H_N ({\cal A})$ are calculated:
\begin{equation}
H_{N} ({\cal A}) = - \sum _{ \lbrace W^{N}({\cal A})\rbrace } 
P(W^{N}({\cal A}))  \ln P(W^{N}({\cal A})).
\end{equation} 
The probabilities $P(W^{N}({\cal A}))$, computed by the frequencies of
$W^{N}({\cal A})$ along a trajectory, are essentially dependent on the
stationary measure selected by the trajectory.  The entropy per unit
time of the trajectory with respect to the partition ${\cal A}$, $h
({\cal A} )$, is defined as follows:
\begin{eqnarray}
h_N({\cal A}) = {1 \over \tau} \lim _{N \to \infty} {1 \over N} H_{N}
({\cal A})\; .
\end{eqnarray} 
Notice that, for the deterministic systems we are considering, the
entropy per unit time does not depend on the sampling time $\tau$
\cite{bill65}. The KS-entropy ($h_{KS}$), by definition, is the
supremum of $h ({\cal A})$ over all possible finite partitions
\cite{bill65,ER85}
\begin{equation}
h_{KS} = \sup _{{\cal A}} h ({\cal A}) \, .
\label{eq:KS}
\end{equation}
The extremal character of $h_{KS}$ makes every computation based on
the definition~(\ref{eq:KS}), impossible in the majority of practical
cases.  In this respect, a useful tool would be the Kolmogorov-Sinai
theorem, through which one is granted that $h_{KS} = h({\cal G})$ if
${\cal G}$ is a generating partition. A partition is said to be
generating if every infinite sequence $\lbrace s_n
\rbrace_{n=1,\dots,\infty} $ corresponds to a single initial
point. However the difficulty now is that, with the exception of very
simple cases, we do not know how to construct a generating
partition. We only know that, according to the Krieger theorem
\cite{krieger}, there exists a generating partition with $k$ elements
such that $e^{h_{KS}} < k \leq e^{h_{KS}} + 1$. Then, a more tractable
way to define $h_{KS}$ is based upon considering the partition ${\cal
A}_{\epsilon}$ made up by a grid of cubic cells of edge $\epsilon$,
from which one has
\begin{equation}
h_{KS} = \lim_{\epsilon \to 0} 
h ({\cal A}_{\epsilon}) \, .
\end{equation}
We expect that $h ({\cal A}_{\epsilon})$ becomes independent of 
$\epsilon$ when ${\cal A}_{\epsilon}$ is so fine to be ``contained''
in a generating partition.

For discrete time maps what has been exposed above is still valid,
with $\tau =1$ (however, Krieger's theorem only applies to invertible
maps).

The important point to note is that, for a truly stochastic
(i.e. non-deterministic) system, with continuous states, $h ({\cal
A}_{\epsilon})$ is not bounded and $ h_{KS} = \infty$.

\subsubsection{Algorithmic complexity}
The Shannon entropy establishes a limit on how efficiently the
ensemble of messages emitted by a source can be coded.  However, we
may wonder about the compressibility properties of a single sequence
with no reference to its belonging to an ensemble.  That is to say, we
are looking for an universal characterization of its compressibility
or, it is the same, an universal definition of its information
content.  This problem can be addressed through the notion of {\em
algorithmic complexity}, that concerns the difficulty in reproducing a
given string of symbols.

Everybody agrees that the binary digits sequence
\begin{equation}
0111010001011001011010...
\label{eq:seq1}
\end{equation}
is, in some sense, more random than 
\begin{equation}
1010101010101010101010...
\label{eq:seq2}
\end{equation}  
The notion of algorithmic complexity, independently introduced by
Kolmogorov \cite{K65}, Chaitin \cite{Ch90} and Solomonov \cite{S64},
is a way to formalize the intuitive idea of randomness of a sequence.

Consider, for instance, a binary digit sequence (this does not
constitute a limitation) of length $N$, $q_N=(i_1,i_2,\dots,i_N)$,
generated by a certain computer code on a given machine ${\cal
M}$. The algorithmic complexity (or algorithmic information content)
$K_{\cal M}(N)$ of $q_N$ is the bit-length of the shortest computer
program able to give $q_N$ and to stop afterward. Of course, such a
length depends not only on the sequence but also on the machine.
However, Kolmogorov \cite{K65} proved the existence of a universal
computer, ${\cal U}$, able to perform the same computation that a
program $p$ makes on ${\cal M}$, with a modification of $p$ that
depends only on ${\cal M}$.  This implies that for all finite strings:
\begin{equation}
K_{\cal U}(N) \leq K_{\cal M}(N)
+ C_{\cal M}\, , 
\label{eq:kolmocomplex}
\end{equation}
where $K_{\cal U}(N)$ is the complexity with respect to the universal
computer and $C_{\cal M}$ depends only on the machine ${\cal M}$.  We
can consider the algorithmic complexity with respect to a universal
computer dropping the ${\cal M}$-dependence in the symbol for the
algorithmic complexity, $K(N)$. The reason is that we are interested
in the limit of very long sequences, $N \to \infty$, for which one
defines the algorithmic complexity per unit symbol:
\begin{equation}
{\cal C}=\lim_{N \to \infty} {K(N) \over N}\,,
\label{eq:acomplexity}
\end{equation}
that, because of (\ref{eq:kolmocomplex}), is an intrinsic quantity,
i.e. independent of the machine.

Now coming back to the $N$-sequences (\ref{eq:seq1}) and
(\ref{eq:seq2}), it is obvious that the latter can be obtained
with a minimal program of length $O(\ln N)$ and
therefore when taking the limit $N \to \infty$ in
(\ref{eq:acomplexity}), one obtains ${\cal C}=0$.  Of course $K(N)$
cannot exceed $N$, since the sequence can always be generated by a
trivial program (of bit length $N$)
\begin{equation}
\nonumber
{\rm "PRINT }\;i_1,i_2,\dots,i_N {\rm "}\,.
\end{equation}
Therefore, in the case of a very irregular sequence, e.g.,
(\ref{eq:seq1}), one expects $K(N) \propto N$ (i.e. ${\cal C}\neq 0$), 
and the sequence is named complex (i.e. of non zero
algorithmic complexity) or random.

Algorithmic complexity cannot be computed, and the un-computability of
$K(N)$ may be understood in terms of G\"odel's incompleteness theorem
\cite{Ch90}. Beyond the problem of whether or not $K(N)$ is computable
in a specific case, the concept of algorithmic complexity brings an
important improvement to clarify the vague and intuitive notion of
randomness.

Between the Shannon entropy, $h_{Sh}$, and the algorithmic complexity, 
there exists the straightforward relationship
\begin{equation}
\lim_{N \to \infty} {\langle K(N) \rangle \over H_N}={1 \over \ln 2}\,,
\label{eq:38}
\end{equation}
where $\langle K(N) \rangle = \sum_{C_N} P(C_N) K_{C_N}(N)$, being
$K_{C_N}(N)$ the algorithmic complexity of the $N$-words, in the ensemble of 
sequences, $C_N$, with a given distribution of probabilities, $P(C_N)$. 
Therefore the expected complexity $\langle K(N) /N\rangle$ is asymptotically
equal to the Shannon entropy (modulo the $\ln 2$ factor).  It is
important to stress again that, apart from the numerical coincidence
of the values of ${\cal C}$ and $h_{Sh}/\ln 2$, there is a conceptual
difference between the information theory and the algorithmic
complexity theory. The Shannon entropy essentially refers to the
information content in a statistical sense, i.e. it refers to an
ensemble of sequences generated by a certain source. The algorithmic
complexity defines the information content of an individual sequence
\cite{G86}. 

The notion of algorithmic complexity can be also applied to the
trajectories of a dynamical system.  This requires the introduction of
finite open coverings of the phase space, the corresponding encoding 
of trajectories into symbolic sequences, and the searching of the supremum 
of the algorithmic complexity per symbol at varying the coverings
\cite{AY81}.  Brudno's and White's theorems \cite{brudno,white} state
that the complexity ${\cal C} ({\bf x})$ for a trajectory starting
from the point ${\bf x}$, is
\begin{equation}
{\cal C} ({\bf x}) = {h_{KS} \over \ln 2} \, ,
\label{eq:dscomplex}
\end{equation}
for almost all ${\bf x}$ with respect to the natural invariant measure.
The factor $\ln 2$ stems again from the conversion between natural
logarithms and bits.

This result indicates that the KS-entropy quantifies not only the
richness of a dynamical system but also the difficulty of describing
its typical sequences.

\subsection{Algorithmic complexity and Lyapunov Exponent}

Let us consider a $1d$ chaotic map 
\begin{equation}
x(t+1)=f(x(t))\,.
\label{eq:mappa}
\end{equation}
The transmission of the sequence $\{ x(t),\;t=1,2,\dots,T\}$,
accepting only errors smaller than a tolerance $\Delta$, is carried
out by using the following strategy \cite{PSV95}:
\begin{enumerate}
\item Transmit the rule 
      (\ref{eq:mappa}): for this task one has to use a number of bits      
      independent of the sequence length $T$.
\item Specify the initial condition $x(0)$ with a precision        
      $\delta_0$ using a finite number of bits which is independent    
      of  $T$.
\item Let the system evolve till the first time $\tau_1$ such that
      the distance between two trajectories, that  was initially
      $\delta x(0)=\delta_0$,
      equals $\Delta$ and then specify again the new   
      initial condition $x(\tau_1)$ with precision $\delta_0$.
\item Let the system evolve and repeat the procedure (2-3), i.e.
      each time the error acceptance tolerance is reached specify
      the initial conditions, 
      $x(\tau_1+\tau_2), \,\;x(\tau_1+\tau_2+\tau_3)\,\dots$, with
      precision $\delta_0$. The times $\tau_1,\tau_2,\dots$
      are defined as follows: putting $x^{'}(\tau_1)=x(\tau_1)+  
      \delta_0$, $\tau_2$ is given by the minimum time such that
      $|x^{'}(\tau_1+\tau_2)-x(\tau_1+\tau_2)| \geq \Delta$ and so  
      on.
\end{enumerate}
Following the steps $(1-4)$, the receiver can reconstruct, with a
precision $\Delta$, the sequence $\{x(t)\}$, by simply iterating on a
computer the evolution law~(\ref{eq:mappa}) between $1$ and $\tau_1-1$,
$\tau_1$ and $\tau_1+\tau_2-1$, and so on. 
The amount of bits necessary to implement the above transmission (1-4)
can be easily computed. 
For simplicity of notation we introduce the quantities
\begin{equation}
\gamma_i={1 \over \tau_i} \ln {\Delta \over \delta_0} \,
\label{eq:mars1}
\end{equation}
which can be regarded as a sort of {\it effective} Lyapunov exponents 
\cite{BPPV85,Fu83}.  
The LE $\lambda$ can be written in terms of $\{\gamma_i\}$ as follows
\begin{equation}
\lambda=\langle \gamma_i \rangle={\sum_i \tau_i \gamma_i \over \sum_i
\tau_i}= {1 \over {\overline {\tau}}} \ln {\Delta \over \delta_0}
\label{eq:liapT}
\end{equation}
where 
$$
{\overline {\tau}} = {1 \over N}\sum \tau_i\,,
$$ is the average time after which we have to transmit the new initial
condition. Note that to obtain $\lambda$ from the
$\gamma_i$'s requires the average (\ref{eq:liapT}), because
the transmission time, $\tau_i$, is not constant.  
If $T$ is large enough the number of transmissions, $N$, is $T/{\overline
{\tau}}\simeq \lambda T/ \ln(\Delta/\delta_0)$.  Therefore, noting
that in each transmission, a reduction of the error from $\Delta$
to $\delta_0$  requires the employ of $\ln_2 (\Delta/\delta_0)$ bits, the
total amount of bits used in the transmission is
\begin{equation}
{T \over {\overline{ \tau}}} \ln_2 {\Delta \over \delta_0}= {\lambda
\over \ln 2} T \,.
\label{eq:bits}
\end{equation}
In other words the number of bits for unit time is proportional to
$\lambda$.

In more than one dimension, we have simply to replace $\lambda$ with
$h_{KS}$ in (\ref{eq:bits}), because the above transmission procedure
has to be repeated for each of the expanding directions.

\section{Limitation of the Lyapunov exponent and Kolmogorov-Sinai
entropy}

Lyapunov exponents and KS-entropy are properly defined only in
specific asymptotic limits: very long times and arbitrary accuracy.
However, predictability problem in realistic situations entails 
considering finite time intervals and limited accuracy.
The first obvious way for quantifying the predictability of a physical
system is in terms of the {\it predictability time} $T_p$, i.e. the
time interval on which one can typically forecast the system.  A
simple argument suggests
\begin{equation}
T_{p} \sim {1 \over \lambda} \ln \left({\Delta \over \delta_0}\right) \,.
\label{eq:2.1-1}
\end{equation}
However, the above relation is too naive to
be of practical relevance, in any realistic system.
Indeed, it does not take into account some
basic features of dynamical systems.  The Lyapunov exponent is a
global quantity, because it measures the average rate of divergence of nearby
trajectories.  In general there exist finite-time fluctuations and their
probability distribution functions (pdf) is
important for the characterization of predictability. The {\it
generalized Lyapunov exponents} have been introduced with the purpose
to take into account such fluctuations \cite{BPPV85,Fu83}.  Moreover,
the Lyapunov exponent is defined for the linearized dynamics, i.e., by
computing the rate of separation of two infinitesimally close
trajectories. On the other hand, in measuring the predictability time
(\ref{eq:2.1-1}) one is interested in a finite tolerance $\Delta$,
because the initial error $\delta_0$ is finite. A recent
generalization of the Lyapunov exponent to {\it finite size} errors
extends the study of the perturbation growth to the nonlinear regime,
i.e. both $\delta_0$ and $\Delta$ are not infinitesimal
\cite{ABCPV96}.

\subsection{Growth of non infinitesimal perturbations}

We discuss now an example where the Lyapunov exponent is of little
relevance for characterizing the predictability.  This problem can be
illustrated by considering the following coupled map model:
\begin{equation}
\left\{\begin{array}{ll} {\bf x}(t+1) &= {\bf R} \, {\bf x}(t) +
\varepsilon {\bf h}(y(t)) \nonumber \\ y(t+1) &= G(y(t)) \nonumber \,,
\end{array}
\right.
\label{eq:2.3-1}
\end{equation}
where ${\bf x} \in {\mathrm I\!R}^{2}$, $y \in {\mathrm I\!R}^{1}$,
${\bf R}$ is a rotation matrix of arbitrary angle $\theta$, ${\bf h}$
is a vector function and $G$ is a chaotic map.  For simplicity we
consider a linear coupling ${\bf h}(y)=(y,y)$ and the logistic map
$G(y)=4 y (1-y)$.

For $\varepsilon=0$ we have two independent systems: a regular and a
chaotic one. Thus the Lyapunov exponent of the ${\bf x}$ subsystem is
$\lambda_{x}(\varepsilon=0)=0$, i.e., it is completely predictable. On
the contrary, the $y$ subsystem is chaotic with
$\lambda_{y}=\lambda_1=\ln 2$.
The switching on of a small coupling ($\varepsilon>0$) yields   
a single three-dimensional chaotic system with a
positive global Lyapunov exponent
\begin{equation}
\lambda = \lambda_{y} + O(\varepsilon) \, .
\label{eq:2.3-2}
\end{equation}
A direct application of (\ref{eq:2.1-1}) would give
\begin{equation}
T_{p}^{(x)} \sim T_{p} \sim {1 \over \lambda_{y}} \, ,
\label{eq:2.3-3}
\end{equation}
but this result is clearly unacceptable: the predictability time for
${\bf x}$ seems to be independent of the value of the coupling
$\varepsilon$. This is not due to an artifact of
the chosen example, indeed, the same argument applies to many
physical situations \cite{BPV96}. A well known example is the
gravitational three body problem, with one body (asteroid) much smaller
than the other two (planets).  When the gravitational
feedback of the asteroid on the two planets is neglected (restricted problem), 
one has a chaotic asteroid in the regular field of the planets. As soon as
the feedback is taken into account (i.e. $\varepsilon>0$ in the
example) one has a non-separable three body system with a positive
LE. 
Of course, intuition correctly suggests that, in the limit of small 
asteroid mass ($\varepsilon \to 0$), a forecast of the planet 
motion should be possible even for very long times.
\begin{figure}
\begin{center}
\includegraphics[scale=.3,angle=-90]{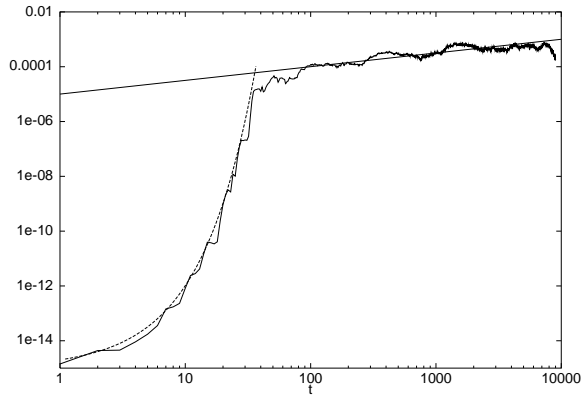}
\end{center}
\caption{\label{fig:2.3-1} Growth of error $|\delta {\bf x}(t)|$ 
for the coupled map
(\ref{eq:2.3-1}).  The rotation angle is $\theta=0.82099$, the
coupling strength $\varepsilon=10^{-5}$ and the initial error only on
the $y$ variable is $\delta y=\delta_0=10^{-10}$. Dashed line $|\delta
{\bf x}(t)|\sim e^{\lambda_1 t}$ where $\lambda_1=\ln 2$, solid line
$|\delta {\bf x}(t)| \sim t^{1/2}$.  }
\end{figure}
The apparent paradox arises from the misuse of formula~(\ref{eq:2.1-1}), 
strictly valid for tangent vectors, to the case of non infinitesimal 
regimes. 
As soon as the errors become large, the full nonlinear evolution of the
three body system has to be taken into account. 
This situation is clearly illustrated by the model (\ref{eq:2.3-1}) in 
Figure~\ref{fig:2.3-1}.  The evolution of
$\delta{\bf x}$ is given by
\begin{equation}
\delta {\bf x}(t+1)={\bf R} \delta {\bf x}(t)+
\varepsilon \delta {\bf h}(y) \, ,
\label{eq:2.3-4}
\end{equation}
where, with our choice, $\delta {\bf h}= (\delta y,\delta y)$.  At the
beginning, both $|\delta {\bf x}|$ and $\delta y$ grow
exponentially. However, the available phase space for $y$ is finite
and the uncertainty reaches the saturation value $\delta y \sim O(1)$
in a time $t \sim 1/\lambda_1$.  At larger times the two realizations
of the $y$ variable are completely uncorrelated and their difference
$\delta y$ in (\ref{eq:2.3-4}) acts as a noisy term. As a consequence,
the growth of the uncertainty on ${\bf x}$ becomes diffusive with a
diffusion coefficient proportional to $\varepsilon^2$ \cite{BPV96}
\begin{equation}
|\delta {\bf x}(t)| \sim \varepsilon t^{1/2} 
\label{eq:2.3-5}
\end{equation}
so that:
\begin{equation}
T_{p}^{(x)} \sim \varepsilon^{-2} \, .
\label{eq:2.3-6}
\end{equation}

This example shows that, even in simple systems, the Lyapunov exponent
can be of little relevance for the characterization of the
predictability.

In more complex systems, in which different scales are present, one is
typically interested in forecasting the large scale motion, while the
LE is related to the small scale dynamics. A familiar example of that is
weather forecast: despite the LE of the atmosphere is indeed rather large, 
due to the small scale convective motion, large-scale weather
predictions are possible for about $10$ days \cite{L69,M73}.  
It is thus
natural to seek for a generalization of the LE to finite perturbations
from which one can obtain a more realistic estimation for the
predictability time.  It is worth underlining the important fact that
finite errors are not confined in the tangent space but are governed
by the complete nonlinear dynamics. In this sense the extension of the
LE to finite errors will give more information on the system.

Aiming to generalize the LE to non infinitesimal perturbations let us
now define the Finite Size Lyapunov Exponent (FSLE) \cite{ABCPV96}.
Consider a reference ${\bf x}(t)$ and a perturbed trajectory ${\bf x}^{'}(t)$, 
such that $|{\bf x}^{'}(0) -{\bf x}(0)| \sim
\delta$.  One integrates the two trajectories and computes the time
$\tau_1(\delta,r)$ necessary for the separation $|{\bf x}^{'}(t)- {\bf
x}(t)|$ to grow from $\delta$ to $r \delta$.  At time
$t\!=\!\tau_1(\delta,r)$ the distance between the trajectories is
rescaled to $\delta$ and the procedure is repeated in order to compute
$\tau_2(\delta,r), \tau_3(\delta,r) \dots$.

The threshold ratio $r$ must be $r>1$, but not too large in order to
avoid contributions from different scales in $\tau(\delta,r)$. A
typical choice is $r=2$ (for which $\tau(\delta,r)$ is properly a
``doubling'' time) or $r=\sqrt{2}$. In the same spirit of the
discussion leading to Eq.s (\ref{eq:mars1}) and (\ref{eq:liapT}), we
may introduce an effective finite size growth rate:
\begin{equation}
\gamma_i(\delta,r)={1 \over \tau_i(\delta,r)}\ln r\,.
\end{equation}

\begin{figure}
\begin{center}
\includegraphics[scale=.7]{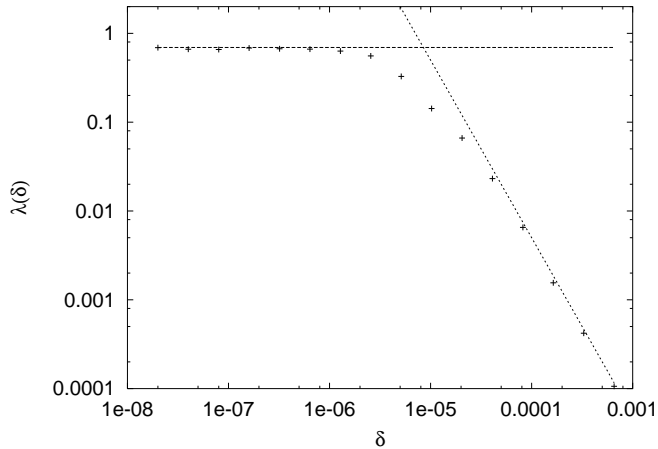}
\end{center}
\caption{\label{fig:2.3-2} $\lambda(\delta)$ as a function of $\delta$
for the coupled map (\ref{eq:2.3-1}) with $\varepsilon=10^{-5}$. The
perturbation has been initialized as in Fig.~\ref{fig:2.3-1}. For
$\delta \to 0$, $\lambda(\delta) \simeq \lambda_1$ (horizontal line). The
dashed line shows the behavior $\lambda(\delta)\sim \delta^{-2}$.}
\end{figure}

After having performed $\cal{N}$ error-doubling experiments, we can
define the FSLE as
\begin{equation}
\lambda(\delta)=\langle\gamma(\delta,r)\rangle_t=
 \left\langle{1 \over
\tau(\delta,r)}\right\rangle_{t} \ln r = {1 \over \langle
\tau(\delta,r) \rangle_{e}} \ln r \, ,
\label{eq:2.3-10}
\end{equation}
where $\langle \tau(\delta,r)\rangle_e$ is
\begin{equation}
\langle \tau(\delta,r)\rangle_{e} =
{1 \over \cal{N}} \sum_{n=1}^{{\cal N}}
\tau_n(\delta,r) \;,
\label{eq:2.3-8}
\end{equation}
see \cite{ABCCV97} for details.  In the infinitesimal limit, the FSLE
reduces to the standard Lyapunov exponent
\begin{equation}
\lim_{\delta \to 0} \lambda(\delta) = \lambda_1 \, .
\label{eq:2.3-11}
\end{equation}
In practice this limit means that $\lambda(\delta)$ displays a
constant plateau at $\lambda_1$ for sufficiently small $\delta$
(Fig.~\ref{fig:2.3-2}). For finite value of $\delta$ the behavior of
$\lambda(\delta)$ depends on the details of the non linear dynamics.
For example, in the model (\ref{eq:2.3-1}) the diffusive behavior
(\ref{eq:2.3-5}), by simple dimensional arguments, corresponds to
$\lambda(\delta)\sim \delta^{-2}$.
Since the FSLE measures the rate of divergence of trajectories at
finite errors, one might wonder whether it is just another way to look
at the average response $\langle \ln(|{\bf x}'(t)-{\bf x}(t)|)
\rangle$ as a function of time. The answer is negative, because 
taking the average at fixed time is not the
same as computing the average doubling time at {\it fixed scale}, as
in (\ref{eq:2.3-10}).  This is particularly clear in the case of
strongly intermittent system, in which $|\delta{\bf x}(t)|$ can be
very different in each realization.  In the presence of intermittency,
averaging over different realizations at fixed times can produce a
spurious regime due to the superposition of exponential and diffusive
contributions by different samples at the same time \cite{ABCCV97}.
The FSLE method can be easily applied to data analysis
\cite{BCPPV98}.  For other approaches addressing the problem of
non-infinitesimal perturbations see \cite{TGP95,KL99}.

\subsection{The $\epsilon$-entropy}
For most systems, the computation of Kolmogorov-Sinai entropy (\ref{eq:KS}) 
is practically impossible, because it involves the limit on arbitrary fine
resolution and infinite times.  However, in the same philosophy of the
FSLE, by relaxing the requirement of arbitrary accuracy, 
one can introduce the $\epsilon$-entropy
which measures the amount of information for reproducing a trajectory
with finite accuracy $\epsilon$ in phase-space.  Roughly speaking the
$\epsilon$-entropy can be considered the counterpart, in information
theory, of the FSLE.  Such a quantity was originally introduced by Shannon
\cite{S48}, and by Kolmogorov \cite{K56}.  Recently Gaspard and Wang
\cite{GW93} made use of this concept to characterize a large variety
of processes.

We start with a continuous-time variable ${\bf x}(t) \in {\mathrm
I\!R}^d$, which represents the state of a $d$-dimensional system, we
discretize the time by introducing an interval $\tau$ and we consider
the new variable
\begin{equation}
\label{eq:2-1}
{\bf X}^{(m)}(t)= \left( {\bf x}(t), {\bf x}(t+\tau), \dots, 
{\bf x}(t+(m-1)\tau) \right). 
\end{equation}
Of course ${\bf X}^{(m)}(t) \in {\mathrm I\!R}^{md}$ and it
corresponds to the trajectory which lasts for a time $T=m \tau$.

In data analysis, the space where the state of the system lives
is  unknown and usually only a scalar variable $u(t)$ can be
measured. Then, one considers vectors $\left( u(t),
u(t+\tau), \dots, u(t+m\tau-\tau) \right)$, that live in ${\mathrm
I\!R}^m$ and allow a reconstruction of the original phase space, known
as delay embedding in the literature \cite{T80,A96,KS97}, and it is a
special case of (\ref{eq:2-1}).
Introduce now a partition of the phase space ${\mathrm I\!R}^d$, using
cells of edge $\epsilon$ in each of the $d$ directions.  Since the
region where a bounded motion evolves contains a finite number of
cells, each ${\bf X}^{(m)}(t)$ can be coded into a word of length $m$,
out of a finite alphabet:
\begin{equation}
\label{eq:2-2}
{\bf X}^{(m)}(t) \longrightarrow W^{m}(\epsilon, t) =
\left( i(\epsilon, t), i(\epsilon, t+\tau), \dots, 
i(\epsilon, t+m \tau -\tau) \right), 
\end{equation}
where $i(\epsilon, t+j \tau)$ labels the cell in ${\mathrm I\!R}^d$
containing ${\bf x}(t+j \tau)$. From the time evolution one obtains,
under the hypothesis of ergodicity, the probabilities
$P(W^{m}(\epsilon))$ of the admissible words $\lbrace W^{m}(\epsilon)
\rbrace$. We can now introduce the $(\epsilon , \tau)$-entropy per
unit time, $h(\epsilon , \tau)$ \cite{S48}:
\begin{equation}
\label{eq:2-3a}
h(\epsilon , \tau) = {1
\over \tau} \lim_{m \to \infty} {1 \over m} H_{m} (\epsilon,\tau) ,
\end{equation} 
where $H_m$ is the block entropy of blocks (words) with length $m$:
\begin{equation}
\label{eq:2-4}
H_{m} (\epsilon,\tau) = - \sum _{ \lbrace W^{m}(\epsilon) \rbrace }
P(W^{m}(\epsilon)) \ln P(W^{m}(\epsilon)).
\end{equation} 
For the sake of simplicity, we ignored the dependence on details of
the partition.  To make $h(\epsilon, \tau)$ partition-independent one
has to consider a generic partition of the phase space $\{{\cal A}\}$
and to evaluate the Shannon entropy on this partition: $h_{Sh}({\cal
A}, \tau)$. The $\varepsilon$-entropy is thus defined as the infimum
over all partitions for which the diameter of each cell is less than
$\varepsilon$ \cite{GW93}:
\begin{equation}
h(\varepsilon, \tau)= \inf_{{\cal A}:{\rm  diam}({\cal A}) \leq
\varepsilon} h_{Sh}({\cal A}, \tau)\,.
\label{def:eps}
\end{equation}
Note that the time dependence in (\ref{def:eps}) is trivial for
deterministic systems, and that in the limit $\epsilon \to 0$ one
recovers the Kolmogorov-Sinai entropy
$$
h_{KS} = \lim_{\epsilon \to 0} h(\epsilon, \tau). 
$$

\section{Characterization of Complexity and system modeling}
In the previous Sections, we discussed the characterization of
dynamical behaviors when the evolution laws are known either exactly
or with some degree of uncertainty.  
In experimental investigations, however, only time records of some 
observable are available, while the equations of motion for the observable 
are generally unknown.  The predictability problem of this
latter case, at least from a conceptual point of view, can be treated
as if the evolution laws were known. Indeed, in
principle, the embedding technique allows for a reconstruction of the phase
space \cite{T80,A96,KS97}. Nevertheless there are rather severe
limitations for high dimensional systems \cite{G89} and even in low
dimensional ones non trivial features appear in the presence of noise
\cite{KS97}.
In this Section we show that an entropic analysis at different
resolution scales provides a pragmatic classification of a signal
and gives suggestions for modeling of systems.  In particular we
illustrate, using some examples, how quantities such as the
$\epsilon$-entropy or the FSLE can display a subtle transition from
the large to the small scales.  A negative consequence of this is the
difficulty in distinguishing, only from data analysis, a genuine
deterministic chaotic system from one with intrinsic randomness
\cite{CFKOV00}.  On the other hand, the way the $\epsilon$-entropy or
FSLE depends on the (resolution) scale,
allows for a classification of the stochastic or chaotic
character of a signal, and this gives some freedom in modeling the
system.

\subsection{How random is a random number generator?}
The ``true character'' of the number sequence $(x_1, x_2, \dots)$ 
obtained by a (pseudo) random number generator (PRNG) on a computer is an 
issue of paramount importance in 
computer simulations and modeling. 
One would like to have a sequence with a random character as much as possible, 
but is forced to use deterministic algorithms to generate 
$(x_1, x_2, \dots)$. This
subsection is mainly based on the paper \cite{KO00}. A simple and
popular PRNG is the multiplicative congruent one:
\begin{equation}
\begin{array}{ll}
z_{n+1} = & N_1 z_n  \:\: \mbox{mod} \:\: N_2  \nonumber \\
x_{n+1} = & z_{n+1}/N_2 \, , 
\end{array}
\label{prng}
\end{equation}
with an integer multiplier $N_1$ and modulus $N_2$. The $\{z_n\}$ are
integer numbers from which one hopes to generate sequence of random
variables $\{x_n\}$, which are uncorrelated and uniformly
distributed in the unit interval.  A first problem arises from
the periodic nature of the rule~(\ref{prng}) as a consequence of its 
discrete nature.  
Note that the rule~(\ref{prng}) can be interpreted also as a
deterministic dynamical system, i.e.
\begin{equation}
x_{n+1} =  N_1 x_n  \:\: \mbox{mod} \:\: 1 \, ,
\label{xdyn}
\end{equation}
which has a uniform invariant measure and a KS entropy $h_{KS}=\lambda
= \ln N_1$.  When imposing the integer arithmetics of Eq.~(\ref{prng})
onto this system, we are, in the language of dynamical systems,
considering an unstable periodic orbit of Eq.~(\ref{xdyn}), with the
particular constraint that, to achieve the period $N_2-1$
(i.e.\ all integers $< N_2$ should belong to the orbit of
Eq.~(\ref{prng})), it has to contain all values $k/N_2$, with
$k=1,2,\cdots, N_2-1$. Since the natural invariant measure of
Eq.~(\ref{xdyn}) is uniform, such an orbit represents the measure of a
chaotic solution in an optimal way.  Every sequence of a PRNG is
characterized by two quantities: its period ${\cal T}$ and its
positive Lyapunov exponent $\lambda$, which is identical to the
entropy of a chaotic orbit of the equivalent dynamical system. Of
course a good random number generator must have a very large period, 
and as large as possible entropy.

\begin{figure}
\begin{center}
\includegraphics[scale=.7]{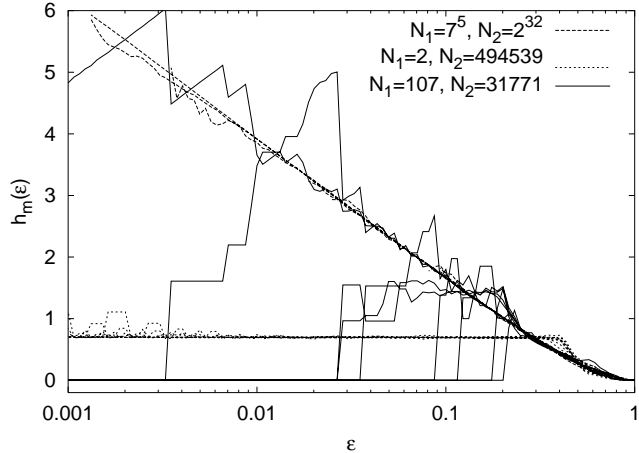}
\end{center}
\caption{\label{fig:rng_entropy} 
The $\epsilon$-entropies, $h_m(\epsilon)$, at varying 
the embedding dimension $m$ for the multiplicative congruential random number
generator Eq.~\ref{prng} for different choices of $N_1$ and $N_2$.}
\end{figure}

It is natural to ask how this apparent randomness can be reconciled
with the facts that (a) the PRNG is a deterministic dynamical systems
(b) it is a discrete state system.
If the period is long enough, on shorter times only
point (a) matters and it can be discussed in terms of the
behavior of the $\epsilon$-entropy, $h(\epsilon)$. 
At high resolutions ($\epsilon \leq 1/N_1$), it seems rather
reasonable to think that the true deterministic chaotic nature of
the congruent rule shows up, and, therefore, $h(\epsilon) \simeq h_{KS}= \ln N_1$. 
On the other hand, for $\epsilon \geq 1/N_1$, one expects to observe the
``apparent random'' behavior of the system, i.e. $h(\epsilon) \sim \ln
(1/ \epsilon)$, see Fig~\ref{fig:rng_entropy}.

\subsection{High dimensional systems}
We discuss an example of high-dimensional system with a non-trivial
behavior at varying the resolution scales, 
namely the emergence of nontrivial collective behavior.

Let us consider a globally coupled map (GCM) defined as follows
\begin{equation}
\label{eq:3.38}
x_n(t+1)=(1-\varepsilon)f_a(x_n(t))+ 
  \frac {\varepsilon} {N} \sum_{i=1}^{N}  f_a(x_{i}(t)),
\end{equation}
where $N$ is the total number of elements, and $f_a(u)$ is a chaotic
map on the interval $[0,1]$, depending on the control parameter $a$.

The evolution of a macroscopic variable, e.g., the center of mass
\begin{equation}
\label{eq:3.39}
m(t)= \frac {1}{N} \sum_{i=1}^{N}  x_{i}(t) ,
\end{equation}
upon varying $\varepsilon$ and $a$ in Eq.~(\ref{eq:3.38}), displays
different behaviors \cite{CFVV99}:
\begin{description}
\item{(a)} {\it Standard Chaos}: $m(t)$ obeys a Gaussian statistics
with a standard deviation $\sigma _{N}=\sqrt{\langle m(t)^2\rangle -
\langle m(t) \rangle ^2} \sim N^{-1/2}$;
\item{(b)} {\it Macroscopic Periodicity}: $m(t)$ 
is a superposition of a periodic function and small fluctuations 
$O(N^{-1/2})$; 
\item{(c)} {\it Macroscopic Chaos}: $m(t)$ exhibits an irregular 
motion, as seen by plotting $m(t)$ vs. $m(t-1)$. The plot 
sketches a structured function (with thickness $\sim N^{-1/2}$), and 
suggests a chaotic motion for $m(t)$. 
\end{description}

In the case of {\it macroscopic chaos}, the center
of mass is expected to evolve with typical times longer than the 
characteristic time $~1/\lambda_1$ of the full dynamics (microscopic dynamics); 
$\lambda_1$ being the Lyapunov exponent of the GCM.
Indeed, conceptually, macroscopic chaos for GCM can be
thought of as the analogous of the hydro-dynamical chaos for molecular
motion. In spite of a huge microscopic Lyapunov exponent
($\lambda_1 \sim 1/\tau_c\sim 10^{11} s^{-1}$, $\tau_c$ is the
collision time), one can have rather different behaviors at a
hydro-dynamical (coarse grained) level: regular motion
($\lambda_{hydro}\leq 0$) or chaotic motion ($0<\lambda_{hydro}\ll
\lambda_1$). In principle, if the hydrodynamic equations were known, a
characterization of the macroscopic behavior would be possible 
by means of standard dynamical system techniques. However, in generic CML there
are no general systematic methods to build up the macroscopic
equations, apart from particular cases \cite{PK94}.  We recall that
for chaotic systems, in the limit of infinitesimal perturbations
$\delta \to 0$, one has $\lambda(\delta) \to \lambda_{1}$, i.e.
$\lambda(\delta)$ displays a plateau at the value $\lambda_{1}$ for
sufficiently small $\delta$. However, for non infinitesimal $\delta$,
one can expect that the $\delta$-dependence of $\lambda(\delta)$ may
give information on the characteristic time-scales governing the
system, and, hence, it could be able to characterize the macroscopic
motion.  In particular, at large scales ($\delta \gg 1/\sqrt{N}$), 
the fast microscopic components saturate and $\lambda(\delta) \approx \lambda_M$, 
where $\lambda_M$ can be fairly called the ``macroscopic'' Lyapunov exponent.

The FSLE has been determined by looking at the evolution of $|\delta
m(t)|$, which has been initialized at the value $\delta
m(t)=\delta_{min}$ by shifting all the elements of the unperturbed
system by the quantity $\delta_{min}$ (i.e. $x^{'}_{i}(0)
=x_{i}(0)+\delta_{min}$), for each realization.  The computation has
been performed by choosing the tent map as local map, but similar
results can be obtained for other maps \cite{SK98,CFVV99}.

The main result can be summarized as follows:
\begin{itemize}
\item at small $\delta\,\,(\ll 1/\sqrt{N})$, where $N$ is the number
of elements, the ``microscopic'' Lyapunov
exponent is recovered, i.e. $\lambda(\delta)\approx
\lambda_{micro}$ 
\item at large $\delta\,\,(\gg 1/\sqrt{N})$, another
plateau $\lambda(\delta) \approx \lambda_{macro}$ appears, 
which can be much smaller than the microscopic one.
\end{itemize}
The emerging scenario is that, at a coarse-grained level,
i.e. $\delta \gg 1/\sqrt{N}$, the system can be described by an
``effective'' hydro-dynamical equation (which in some cases can be
low-dimensional), while the ``true'' high-dimensional character
appears only at very high resolution, i.e.
$$
\delta \leq \delta_c = O\left({1 \over \sqrt{N}}\right).
$$

\subsection{Diffusion in deterministic  systems and Brownian motion}
Consider the following map which generates a diffusive behavior on the
large scales \cite{SFK82}:
\begin{equation}
\label{eq:3-1}
x_{t+1} = \lbrack x_{t} \rbrack + F\left(x_{t} - \lbrack x_{t} \rbrack
 \right) ,
\end{equation} 
where $\lbrack x_{t} \rbrack$ indicates the integer part of $x_{t}$
and $F(y)$ is given by:
\begin{equation}
F(y)=\left\{\begin{array}{ll}
(2+\alpha) y & \:\: \mbox{if}\:\: y\, \in [0,1/2[ \nonumber \\
(2+\alpha) y-(1+\alpha) &\:\: \mbox{if}\:\: y\, \in \,[1/2,1]\,.
\end{array}\right.
\label{eq:mappaF}
\end{equation}
The largest Lyapunov exponent $\lambda$ can be obtained immediately:
$\lambda = \ln |F'|$, with $F'=dF/dy=\!2\!+\!\alpha$.  One expects the
following scenario for $h(\epsilon)$:
\begin{equation}
\label{eq:3-2}
h(\epsilon) \approx \lambda \qquad {\rm for} \quad \epsilon < 1 ,
\end{equation} 
\begin{equation}
\label{eq:3-3}
h(\epsilon) \propto {D\over \epsilon ^2} \qquad {\rm for} 
\quad \epsilon > 1 , 
\end{equation} 
where $D$ is the diffusion coefficient,
$\langle \left( x_t -x_0 \right)^2 \rangle \approx 2 \,\, D \,\, t
\qquad {\rm for} \quad {\rm large} \quad t $.
\begin{figure}
\begin{center}
\includegraphics[scale=.7]{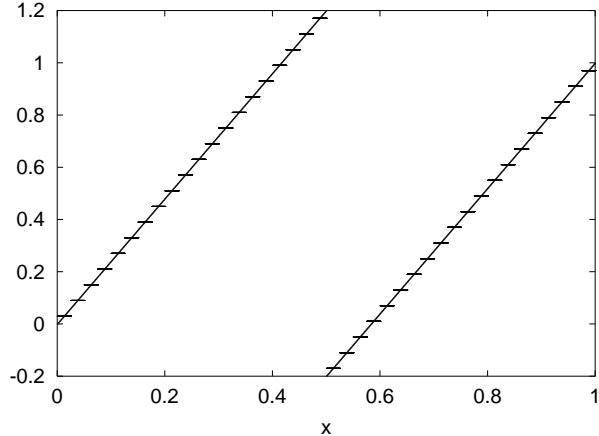}
\end{center}
\caption{\label{map} The map $F(x)$ (\ref{eq:mappaF}) for $\alpha=0.4$ 
is shown
with superimposed the approximating (regular) map $G(x)$
(\ref{eq:3-5}) obtained by using $40$ intervals of slope $0$. }
\end{figure}
Consider now a stochastic system, namely a noisy map
\begin{equation}
\label{eq:3-5}
x_{t+1} = \lbrack x_{t} \rbrack + G \left(x_{t} - \lbrack x_{t} \rbrack
 \right) + \sigma \eta _{t},
\end{equation} 
where $G(y)$, as shown in Fig.~\ref{map}, is a piece wise linear map
which approximates the map $F(y)$, and $\eta _{t}$ is a stochastic
process uniformly distributed in the interval $\lbrack -1, 1 \rbrack$,
and no correlation in time.  When $|dG/dy| < 1$, as is the case we
consider, the map (\ref{eq:3-5}), in the absence of noise, gives a
non-chaotic time evolution.
\begin{figure}
\begin{center}
\includegraphics[scale=.7]{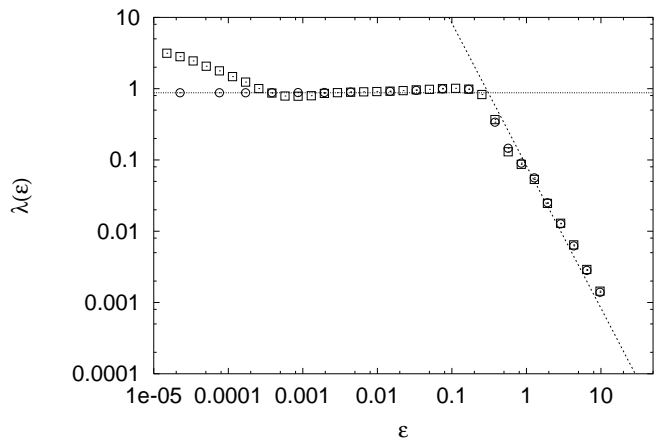}
\end{center}
\caption{\label{fslediff} 
Lyapunov exponent $\lambda(\epsilon)$ versus 
$\epsilon$ obtained for the map
$F(y)$ (\ref{eq:mappaF}) with $\alpha=0.4$ ($\circ$) and for the noisy
(regular) map (\ref{eq:3-5}) ($\Box$)  with $10^4$ intervals of slope
$0.9$ and $\sigma=10^{-4}$. Straight lines indicate the Lyapunov
exponent $\lambda=\ln 2.4$ and the diffusive behavior
$\lambda(\epsilon) \sim \epsilon^{-2}$}.
\end{figure}

Now we compare the finite size Lyapunov exponent for the chaotic map
(\ref{eq:3-1}) and for the noisy one (\ref{eq:3-5}). In the latter the
FSLE has been computed using two different realizations of the noise.
In Fig.~\ref{fslediff} we show $\lambda(\epsilon)$ versus $\epsilon$
for the two cases.  The two curves are practically indistinguishable
in the region $\epsilon >\sigma$. The differences appear only at very
small scales $\epsilon < \sigma$ where one has a $\lambda(\epsilon)$
which grows with $\epsilon$ for the noisy case, remaining at the same
value for the chaotic deterministic case.

Both the FSLE and the $(\epsilon,\tau)$-entropy analysis show that we
can distinguish three different regimes observing the dynamics of
(\ref{eq:3-5}) on different length scales. On the large length scales
$\epsilon > 1$ we observe diffusive behavior in both models. On length
scales $\sigma < \epsilon < 1$ both models show chaotic deterministic
behavior, because the entropy and the FSLE are independent of
$\epsilon$ and larger than zero. Finally on the smallest length scales
$\epsilon < \sigma $ we see stochastic behavior for the system
(\ref{eq:3-5}), {\em i.e.} $h(\varepsilon) \sim -\ln(\varepsilon)$,
while the system (\ref{eq:3-1}) still shows chaotic behavior.

\subsection{On the distinction between chaos and noise}
\label{sec:7.2.4}

The above examples show that the distinction between chaos and noise
can be a highly non trivial task, which makes sense only in very
peculiar cases, e.g., very low dimensional systems.  Nevertheless,
even in this case, the entropic analysis can be unable to recognize
the ``true'' character of the system due to the lack of
resolution. Again, the comparison between the diffusive map
(\ref{eq:3-1}) and the noisy map (\ref{eq:3-5}) is an example of these
difficulties.  For $\sigma \leq \epsilon \leq 1$ both the system
(\ref{eq:3-1}) and (\ref{eq:3-5}), in spite of their ``true''
character, will be classified as chaotic, while for $\epsilon \geq 1$
both can be considered as stochastic.

In high-dimensional chaotic systems, with $N$ degrees of freedom, one
has typically $h(\epsilon)=h_{KS}\sim O(N)$ for $\epsilon \leq
\epsilon_c$ (where $\epsilon_c \to 0$ as $N\to \infty$) while for
$\epsilon\geq \epsilon_c$, $h(\epsilon)$ decreases, often with a power
law \cite{GW93}.  Since also in some stochastic processes the
$\epsilon$-entropy obeys a power law, this can be a source of
confusion.

These kind of problems are not abstract ones, as a recent debate on
``microscopic chaos'' demonstrates \cite{GBFSGDC98,DCB99,GS99}.  The
detection of microscopic chaos by data analysis has been recently
addressed in a work of Gaspard et al.~\cite{GBFSGDC98}.  These
authors, from an entropic analysis of an ingenious experiment on the
position of a Brownian particle in a liquid, claim to give an
empirical evidence for microscopic chaos.  In other words, they state
that the diffusive behavior observed for a Brownian particle is the
consequence of chaos at a molecular level.  Their work can be briefly
summarized as follows: from a long ($\approx 1.5 \times 10^5$ data)
record of the position of a Brownian particle they compute the
$\epsilon$-entropy with the Cohen-Procaccia method \cite{CP85} from
which they obtain:
\begin{equation} 
h(\epsilon) \sim {D \over \epsilon^2}\,,
\label{eq:gasp}
\end{equation} 
where $D$ is the diffusion coefficient. Then, {\em assuming} that the
system is deterministic, and making use of the inequality $h(\epsilon
>0) \leq h_{KS}$, they conclude that the system is chaotic. However,
their result does not give a direct evidence that the system is
deterministic and chaotic.  Indeed, the power law (\ref{eq:gasp}) can
be produced with different mechanisms:
\begin{enumerate}
\item a genuine chaotic system with diffusive behavior, 
      as the map (\ref{eq:mappaF});
\item a non chaotic system with some noise, as the map (\ref{eq:3-5}), 
     or a genuine Brownian system;
\item a deterministic linear non chaotic system with many degrees of
freedom (see for instance \cite{MM60});
\item a ``complicated'' non chaotic system as the Ehrenfest wind-tree
      model where a particle diffuses in a plane due to collisions with
      randomly placed, fixed oriented square scatters, as discussed by Cohen
      et al. \cite{DCB99} in their comment to Ref.~\cite{GBFSGDC98}.
\end{enumerate}
It seems to us that the weak points of the analysis in
Ref.~\cite{GBFSGDC98} are:
\begin{description}
\item{a)} the explicit assumption that the system is deterministic;
\item{b)} the limited number of data points and therefore 
limitations in both resolution and block length.
\end{description}
The point (a) is crucial, without this assumption (even with an
enormous data set) it is not possible to distinguish between 1) and
2).  One has to say that in the cases 3) and 4) at least in principle
it is possible to understand that the systems are ``trivial''
(i.e. not chaotic) but for this one has to use a huge number of
data. For example Cohen et al.~\cite{DCB99} estimated that in order to
distinguish between 1) and 4) using realistic parameters of a typical
liquid, the number of data points required has to be at least $\sim
10^{34}$.

Concluding, we have the apparently paradoxical result that
``complexity'' helps in the construction of models. Basically, in the
case in which one has a variety of behaviors at varying the scale
resolution, there is a certain freedom on the choice of the model to
adopt.  For some systems the behavior at large scales can be realized
both with chaotic deterministic models or suitable stochastic
processes. From a pragmatic point of view, the fact that in certain
stochastic processes $h(\epsilon) \sim \epsilon^{-\alpha}$ can be
indeed extremely useful for modeling such high-dimensional
systems. Perhaps, the most relevant case in which one can use this
freedom in modeling is the fully developed turbulence whose non
infinitesimal (the so-called inertial range) properties can be
successfully mimicked in terms of multi-affine stochastic process (see
Ref.~\cite{BBCCV98}).

\section{Concluding Remarks}
The guideline of this paper has been {\em the interpretation of 
different aspects of the predictability of a system as a way to 
characterize its complexity}.

We have discussed the relation between chaoticity, the
Kolmogorov-Sinai entropy and algorithmic complexity. As clearly
exposed in the seminal works of Alekseev and Yakobson \cite{AY81} and
Ford \cite{F83}, the time sequences generated by a system with
sensitive dependence on initial conditions have non-zero algorithmic
complexity. A relation exists between the maximal compression of a
sequence and its KS-entropy.  Therefore, one can give a definition of
complexity, without referring to a specific description, as an
intrinsic property of the system.

The study of these different aspects of predictability constitutes a
useful method for a quantitative characterization of ``complexity'',
suggesting the following equivalences:
\begin{equation}
{\bf Complex} = {\bf Uncompressible} = {\bf Unpredictable}
\end{equation}
The above point of view, based on dynamical systems and information
theory, quantifies the complexity of a sequence considering each
symbol relevant but it does not capture the structural level.  Let us
clarify this point with the following example.  A binary sequence
obtained with a coin tossing is, from the point of view adopted in
this review, complex since it cannot be compressed (i.e. it is
unpredictable). On the other hand such a sequence is somehow trivial,
i.e.  with low ``organizational'' complexity. It would be important to
introduce a quantitative measure of this intuitive idea. The
progresses of the research on this intriguing and difficult issue are
still rather slow.  We just mention some of the most promising
proposals as the logical depth and the sophistication~\cite{BP97}.

\hfill\break

\noindent {\bf Acknowledgments} We thank G.~Boffetta, A.~Celani, 
D.~Vergni and M.~Cencini for the long collaboration and many fruitful 
discussions on the subject of the paper.



\begin{thebibliography}{99}
\bibitem {L1814} 
S.~Laplace, 
{\it Essai philosophique sur les probabilit\'es}, 
Courcier, Paris  (1814).

\bibitem{Boffy} G.~Boffetta, M.~Cencini, M.~Falcioni and A.~Vulpiani  
``Predictability: a way to characterize complexity'', 
Phys. Rep. {\bf 356} (2002) 367. 

\bibitem{O68} 
V.I.~Oseledec, 
``A multiplicative ergodic theorem: 
Lyapunov characteristic numbers for dynamical systems'', 
Trans. Mosc. Math. Soc.  {\bf 19} (1968) 197. 

\bibitem{S48} 
C.E.~Shannon, 
``A mathematical theory of communication'' 
{\it The Bell System Technical J.} 
{\bf 27} (1948) 623; {\bf 27} (1948) 379.

\bibitem{K65} 
A.N.~Kolmogorov, 
``Three approaches to the quantitative definition of information'', 
Prob. Info. Trans. {\bf 1} (1965) 1.

\bibitem {S64} 
R.J.~Solomonoff, 
``A formal theory of inductive inference'', 
Inform. Contr. {\bf 7} (1964)  1; {\bf 7}  (1964) 224.  

\bibitem{K57} 
A.I.~Khinchin, 
{\it Mathematical foundations of information theory}, 
Dover, New York  (1957).

\bibitem{N1}
D.~Welsh, {\it Codes and Cryptography},
Clarendon Press, Oxford 1989. 

\bibitem{bill65} 
P.~Billingsley, 
{\it Ergodic theory and information},
Wiley, New York (1965).

\bibitem {ER85} 
J.P.~Eckmann and D.~Ruelle, 
``Ergodic theory of chaos and strange attractors'', 
Rev. Mod. Phys. {\bf 57} (1985) 617.

\bibitem{krieger}
W.~Krieger, 
``On entropy and generators of measure preserving transformations '', 
Trans. Am. Math. Soc. {\bf 149} (1970) 453.

\bibitem {Ch90} 
G.J.~Chaitin, 
{\it Information, randomness and incompleteness } 
2nd edition, World Scientific, Singapore (1990).

\bibitem{G86} 
P.~Grassberger, ``Toward a quantitative theory of
self-generated complexity'', Int. J. Theor. Phys. {\bf 25} (1986) 907.

\bibitem{AY81} 
 V.M.~Alekseev and M.V.~Yakobson, ``Symbolic dynamics and
hyperbolic dynamic-systems'', 
Phys. Rep. {\bf 75} (1981) 287. 

\bibitem{brudno} 
A.A.~Brudno, 
``Entropy and the complexity of the trajectories of a dynamical system'', 
Trans. Moscow Math. Soc. {\bf 44} (1983) 127. 

\bibitem{white} 
H.~White, 
``Algorithmic complexity of Points in  Dynamical Systems'', 
Erg. Theory Dyn. Syst. {\bf 13} (1993) 807. 

\bibitem{F83} 
J.~Ford, ``How Random is a Coin Tossing?'', 
Physics Today, {\bf 36} (1983) 40.
 
\bibitem {PSV95}
G.~Paladin, M.~Serva and A.~Vulpiani, 
``Complexity in dynamical systems with noise'', 
Phys. Rev. Lett. {\bf 74} (1995) 66.

\bibitem{Fu83} 
H.~Fujisaka, 
``Statistical dynamics generated by fluctuations of local 
Lyapunov exponents'', 
Prog. Theor. Phys. {\bf 70} (1983) 1264. 

\bibitem{BPPV85} 
R.~Benzi, G.~Paladin, G.~Parisi and A.~Vulpiani,
``Characterization of intermittency in chaotic systems'', 
J. Phys. A {\bf 18} (1985) 2157. 

\bibitem{ABCPV96} 
E.~Aurell, G.~Boffetta, A.~Crisanti, G.~Paladin and A.~Vulpiani, 
``Growth of Non-infinitesimal Perturbations in Turbulence'', 
Phys. Rev. Lett. {\bf 77} (1996) 1262.

\bibitem{BPV96} 
G.~Boffetta, G.~Paladin and A.~Vulpiani, 
``Strong Chaos without Butterfly Effect 
in Dynamical Systems with Feedback'',  
J. Phys. A, {\bf 29} (1996) 2291.  

\bibitem{L69} 
E.N.~Lorenz, 
``The predictability of a flow which possesses many scales of motion'', 
Tellus, {\bf 21} (1969) 3. 

\bibitem{M73} 
A.~Monin, 
{\it Weather prediction as a problem in physics}, 
MIT Press, Moscow (1973).

\bibitem{ABCCV97} 
V.~Artale, G.~Boffetta, A.~Celani, M.~Cencini and A.~Vulpiani, 
``Dispersion of passive tracers in closed basins: Beyond the 
diffusion coefficient'',  
Phys. Fluids A, {\bf 9} (1997) 3162.

\bibitem{BCPPV98}  
G. Boffetta, A. Crisanti, F. Paparella, A. Provenzale  and A. Vulpiani,  
``Slow and fast dynamics in coupled systems:  a time series analysis view'', 
Physica D, {\bf 116} (1998) 301. 

\bibitem{KL99} 
H.~Kantz and T.~Letz, 
``Quasi-chaos and quasi-regularity- the breakdown of linear 
stability analysis'', 
Phys. Rev. E, {\bf 61} (2000) 2533.

\bibitem{TGP95} 
A.~Torcini, P.~Grassberger and A.~Politi, 
``Error propagation in extended chaotic systems'', 
J. Phys. A, {\bf 27} (1995) 4533. 

\bibitem {K56} 
A.N.~Kolmogorov, 
``On the Shannon theory of information transmission 
in the case of continuous signals'', 
IRE Trans. Inf. Theory, {\bf 1} (1956) 102.

\bibitem{GW93} 
P.~Gaspard and X.J.~Wang, 
``Noise, chaos, and $(\epsilon,\tau)$-entropy per unit time'', 
Phys. Rep. {\bf 235} (1993) 291. 

\bibitem{T80}
F.~Takens,  
``Detecting strange attractors in turbulence''
in {\em Dynamical Systems and Turbulence (Warwick 1980)}, 
Vol.~898 of {\em Lecture Notes in Mathematics}, D.A.~Rand and L.-S.~Young
(eds.), pg. 366, Springer-Verlag, Berlin (1980).

\bibitem {KS97} 
H.~Kantz and T.~Schreiber, 
{\it Nonlinear time series analysis}, 
Cambridge University Press, Cambridge  (1997).

\bibitem{A96}
 H.D.I.~Abarbanel, 
{\it Analysis of Observed Chaotic Data},
Springer-Verlag, New York (1996).

\bibitem {G89} 
P.~Grassberger, 
``Information content and predictability of lumped and distributed 
dynamical systems'', 
Phisica Scripta, {\bf 40} (1989) 346.

\bibitem{CFKOV00}
M.~Cencini, M.~Falcioni, H.~Kantz, E.~Olbrich and A.~Vulpiani, 
``Chaos or Noise --- Sense and Nonsense of a Distinction'', 
Phys. Rev. E, {\bf 62} (2000) 427. 

\bibitem {KO00} 
H.~Kantz and E.~Olbrich, 
``Coarse grained dynamical entropies: investigation of high 
entropic dynamical systems'',
Physica A,  {\bf 280} (2000) 34. 

\bibitem {PK94} 
A.S.~Pikovsky and J.~Kurths, 
``Collective behavior in ensembles of globally coupled maps'', 
Physica D, {\bf 76} (1994) 411.

\bibitem{CFVV99} 
M.~Cencini, M.~Falcioni, D.~Vergni and A.~Vulpiani,
``Macroscopic chaos in globally coupled maps'', 
Physica D, {\bf 130} (1999) 58.

\bibitem{SK98} 
T.~Shibata and K.~Kaneko. 
``Collective Chaos'', 
Phys. Rev. Lett. {\bf 81} (1998) 4116. 

\bibitem{SFK82}
M.~Schell, S.~Fraser and R.~Kapral, 
``Diffusive dynamics in systems with translational 
symmetry --A one-dimensional-map model'',
Phys. Rev. A, {\bf 26} (1982)  504.  

\bibitem{GBFSGDC98}
P.~Gaspard, M.E.~Briggs, M.K.~Francis, J.V.~Sengers,
R.W.~Gammon, J.R.~Dorfman and R.V.~Calabrese, 
``Experimental evidence for microscopic chaos'',
Nature {\bf 394} (1998) 865.  

\bibitem{DCB99}
C.~Dettman, E.~Cohen, and H.~van~Beijeren, 
``Microscopic chaos from Brownian motion?'' 
Nature {\bf 401} (1999)  875.  

\bibitem{GS99}
P. Grassbeger and T. Schreiber, 
``Statistical mechanics - Microscopic chaos from Brownian motion?'',
Nature {\bf 401} (1999) 875.

\bibitem{CP85} 
A.~Cohen and I.~Procaccia, 
``Computing the Kolmogorov entropy from time signals of dissipative
and conservative dynamical systems'', 
Phys. Rev. A, {\bf 31} (1985) 1872.

\bibitem{MM60}
P.~Mazur and E.~Montroll, 
``Poincare cycles, ergodicity and irreversibility in assemblies 
of coupled harmonic oscillators'',
J. Math. Phys. {\bf 1} (1960) 70.  

\bibitem{BBCCV98} 
L. Biferale, G. Boffetta, A. Celani, A. Crisanti and A. Vulpiani,
``Mimicking a turbulent signal: sequential multi-affine processes'',
Phys. Rev. E, {\bf 57} (1998) R6261.

\bibitem{BP97} 
R.~Badii and A.~Politi, 
{\it Complexity. Hierarchical structures and scaling in physics}, 
Cambridge University Press, Cambridge, UK  (1997).

\end{thebibliography}
\end{document}